\documentclass{svproc} 
\usepackage{float}
\usepackage{newtxtext}
\usepackage{newtxmath}
\usepackage{tabularx}
\usepackage{multirow}
\usepackage{adjustbox}
\usepackage{booktabs} 
\usepackage{ragged2e} 
\usepackage{longtable} 
\usepackage{enumitem} 
\usepackage{float}

\usepackage{array}

\usepackage[T1]{fontenc}
\usepackage{indentfirst}
\usepackage{wrapfig}
\usepackage{textcomp}

\usepackage{cite}

\usepackage{algorithm}
\usepackage{algpseudocode}
\usepackage{amsmath}

\usepackage{amsthm}
\usepackage{lipsum} 
\usepackage[english]{babel}
\usepackage[autostyle]{csquotes}

\title{Securing RC Based P2P Networks: A Blockchain-based Access Control Framework utilizing Ethereum Smart Contracts for IoT and Web 3.0 
\thanks{Corresponding author: rmitra@semo.edu}}

\author{
Saurav Ghosh\inst{1}, Reshmi Mitra\inst{1}, Indranil Roy\inst{1}, Bidyut Gupta\inst{2}
}

\institute{
  Southeast Missouri State University, Cape Girardeau, USA \\
  \email{\{sghosh3s, rmitra, iroy\}@semo.edu}
  \and
  Southern Illinois University, Carbondale, USA \\
  \email{bidyut@cs.siu.edu}
}

\authorrunning{Ghosh, Mitra, and Roy}
\titlerunning{Blockchain-based Access Control for P2P Networks}

\begin{document}

\maketitle

\begin{abstract}
Ensuring security for highly dynamic peer-to-peer (P2P) networks has always been a challenge, especially for services like online transactions and smart devices. These networks experience high churn rates, making it difficult to maintain appropriate access control. Traditional systems, particularly Role-Based Access Control (RBAC), often fail to meet the needs of a P2P environment. This paper presents a blockchain-based access control framework that uses Ethereum smart contracts to address these challenges. Our framework aims to close the gaps in existing access control systems by providing flexible, transparent, and decentralized security solutions. The proposed framework includes access control contracts (ACC) that manage access based on static and dynamic policies, a Judge Contract (JC) to handle misbehavior, and a Register Contract (RC) to record and manage the interactions between ACCs and JC. The security model combines impact and severity-based threat assessments using the CIA (Confidentiality, Integrity, Availability) and STRIDE principles, ensuring responses are tailored to different threat levels. This system not only stabilizes the fundamental issues of peer membership but also offers a scalable solution, particularly valuable in areas such as the Internet of Things (IoT) and Web 3.0 technologies. 
\end{abstract}

\section{Introduction} \label{Sec:introduction}

Studying security issues in peer-to-peer (P2P) networks is crucial due to their inherent vulnerabilities arising from decentralized control, which makes them prime targets for various attacks such as Sybil, man-in-the-middle, and denial-of-service (DoS) attacks. In P2P networks, each peer acts as both a client and server, creating numerous points of potential failure and compromise. Traditional role-based access control (RBAC) models are inadequate because they are designed for centralized systems where roles and permissions are clearly defined and controlled by a central authority. The lack of centralization in P2P networks makes it difficult to enforce consistent access controls and manage identities securely.

One significant issue with RBAC in P2P networks is Role Explosion, where managing multiple roles for all possible entities becomes unmanageable. The static nature of roles in RBAC conflicts with the dynamic characteristics of P2P networks, where nodes frequently join and leave. This constant change leads to an explosion of roles, complicating the maintenance and updating of role definitions, and creating potential security gaps. Addressing security in P2P networks requires developing more robust and adaptable security mechanisms capable of handling their dynamic and decentralized nature, managing identities and access controls in real time, and overcoming the limitations of traditional RBAC models.

In this paper, we introduce a blockchain-based access control architecture that tackles these challenges by harnessing the properties of blockchain: constant verification, immutability, transparency, and decentralization. Blockchain's ongoing verification processes ensure that each transaction and access request is validated by multiple nodes, creating a resilient security framework suitable for dynamic and unsecured environments such as P2P networks \cite{islam2019permissioned} \cite{sun2019blockchain}. Our proposed architecture consists of three main components:

\begin{itemize}[noitemsep, topsep=0pt]
\item \textbf{Register Contract (RC):} This maintains a comprehensive registry of all access control methods and their corresponding smart contracts, making it easier to register, update, and delete methods. The concept of an RC was initially proposed by Zhang et al. \cite{zhang2018smart}. Our work expands on it by incorporating dynamic policy management and real-time updates to better handle the fluid nature of P2P networks.
\item \textbf{Judge Contract (JC):} This contract evaluates reported misbehaviors, determines penalties, and ensures appropriate actions are taken against violators, guaranteeing dynamic and fair adjudication of security breaches.
\item \textbf{Access Control Contracts (ACCs):} These contracts implement specific access control policies for subject-object pairs. They perform static validation based on predefined policies and dynamic validation by monitoring subject behavior.
\end{itemize}
Our work stands out for its comprehensive approach to integrating blockchain technology with access control mechanisms. Unlike previous studies that mainly used blockchain for storing access policies or managing data records \cite{islam2019permissioned} \cite{sun2019blockchain}, our framework leverages the computational capabilities of smart contracts to enforce security policies dynamically. This dual-layer approach significantly enhances security, scalability, and adaptability in decentralized networks. The main contributions of this paper are:
\begin{itemize}[noitemsep, topsep=0pt]
\item \textbf{Blockchain-Based Access Control Framework:} We developed a new access control framework using Ethereum smart contracts to improve security in decentralized networks.
\item \textbf{Policy Management:} We introduced mechanisms for effective policy management within the framework, capable of adapting in real-time to network conditions and security threats.
\item \textbf{Scalable Security Solutions for Decentralized Networks:} We introduced a blockchain-based platform design for IoT and Web 3.0 solutions, allowing secure decentralized system growth.
\item \textbf{Empirical Validation and Practical Application:} Demonstrated usefulness through practical application, showcasing potential for large-scale adoption.
\end{itemize}
This paper has six main sections. In Section~\ref{Sec:related_work}, the summary of the access control and blockchain applications in P2P networks is presented. Section~\ref{sect:Prelims} describes the RC-based P2P architecture. Section~\ref{Sec:methods} describes the design, whereas Section~\ref{Sec:implementation} describes the implementation of the blockchain-based access control framework, including the integration of the Ethereum smart contracts. Section~\ref{Sec:result} discusses the results and evaluation of the framework's impact on network security through several deployments. Finally, Section~\ref{Sec:conclusion} concludes with the accomplishments of this research.

\section{Related Work} \label{Sec:related_work}

\textbf{Security in IoT and P2P Networks}: The need for robust security frameworks in IoT environments is critical. Novo \cite{novo2018blockchain} proposed scalable blockchain architecture for IoT access management, demonstrating how blockchain can enhance security and operational efficiency. Yu, Chen, and Wang \cite{yu2022blockchain} highlighted additional security challenges, such as latency and node heterogeneity in distributed networks. Mbarek, Ge, and Pitner \cite{mbarek2020blockchain} extended these concepts to smart home and smart grid systems, emphasizing the improved data integrity and resilience against cyber-attacks within peer-to-peer networks.

\textbf{Blockchain-Based Access Control Mechanisms}: Sun et al. \cite{sun2022blockchain} reviewed the transformative role of blockchain in IoT access control, focusing on enhanced transparency through smart contracts. Ding et al. \cite{ding2019attribute} proposed an attribute-based access control scheme that minimizes overhead while enhancing security and practicality. Similarly, Wang et al. \cite{wang2022dynamic} developed a dynamic access control system that adapts to user behavior and device status in real-time, showcasing blockchain's flexibility in complex network environments.
%
Liu et al. \cite{liu2020fabric} demonstrated the impact of integrating blockchain into IoT networks to enhance access control mechanisms, thereby reducing risks from unauthorized access and breaches. Practical examples provided by Sultana et al. \cite{sultana2020data} and Badhe and Arjunwadkar \cite{badhe2023access} illustrated the scalability and adaptability of blockchain-based smart contracts for secure data sharing and distribution in contemporary IoT frameworks.


Collaborating access control systems with blockchain models in P2P networks offers a robust solution for IoT security. Zhang et al. \cite{zhang2020review} provided a critical evaluation of various blockchain-based models, identifying key areas for scalability and security improvements. Han et al. \cite{han2022blockchain} and Li et al. \cite{li2021iot} further elaborated on enhancing access control through dual-layer blockchain systems, promoting reliability and flexibility. 

\textit{Our work builds on these foundations by integrating dynamic policy management and real-time updates}. This means our approach can easily adapt to the ever-changing nature of P2P networks. This advancement is a game-changer. It shows just how practical and applicable our proposed framework is on a large scale, making a significant contribution to the field of decentralized network security. Our framework goes above and beyond by tackling the limitations of existing models and boosting the overall security and operational integrity of decentralized environments.

\section{Preliminaries} \label{Sec:prelim}

\label{sect:Prelims}
    We have taken into consideration some of the first results of an RC-based low-diameter two-level hierarchical structured P2P network \cite{neupane2023efficient}, which is a two-level overlay architecture, and at each level structured networks of peers exist. It is explained in detail below.
 \begin{enumerate}[noitemsep, topsep=0pt]
    \item At level-1, we have a ring network consisting of the peers $H_{i}$ $(0 \leq i \leq n-1)$. The number of peers on the ring is n, the number of distinct resource types. This ring network is used for efficient data lookup, so we name it a transit ring network.
    \item At level 2, there are n numbers of completely connected networks (groups) of peers. Each such group, say $G_{i}$ is formed by the peers of the subset $P^{Ri}$, $(0 \leq i \leq n-1)$, such that all peers $(\in P^{Ri})$ are directly connected (logically) to each other, resulting in the network diameter of 1. Each $G_{i}$ is connected to the transit ring network via its group-head $H_{i}$.
    \item Each peer in the network maintains an Information Resource Table (IRT) that consists of n number of tuples.
\end{enumerate}

\section{Design of Access Control Framework} \label{Sec:methods}

Blockchain-based access control distributes the validation and enforcement of policies across multiple nodes in the network. This means that access requests and transactions are continuously verified by multiple participants, which ensures security and trust among network members and eliminates the risk of a single point of failure.
The proposed framework consists of multiple access control contracts (ACCs), a Judge Contract (JC), and a Register Contract (RC). Each network entity, such as group heads and members, is assigned access privileges based on roles and responsibilities. These privileges are managed at a granular level by various ACCs, while the RC \cite{islam2019permissioned} \cite{sun2019blockchain} \cite{dhif2019blockchain} maintains a system-level view of all access control policies. The JC receives regular monitoring updates from ACCs about various STRIDE events and issues penalties accordingly.
Our design philosophy focuses on \textit{flexibility, scalability, and robustness}. The architecture dynamically manages access permissions and enforces security policies in real time, accommodating high churn rates and the structural openness of P2P networks. Using Ethereum smart contracts, our framework can implement complex access control policies and respond effectively to security threats.

\begin{figure}[ht]
    \centering
    \begin{minipage}{0.5\textwidth}
        \centering
        \includegraphics[width=\textwidth]{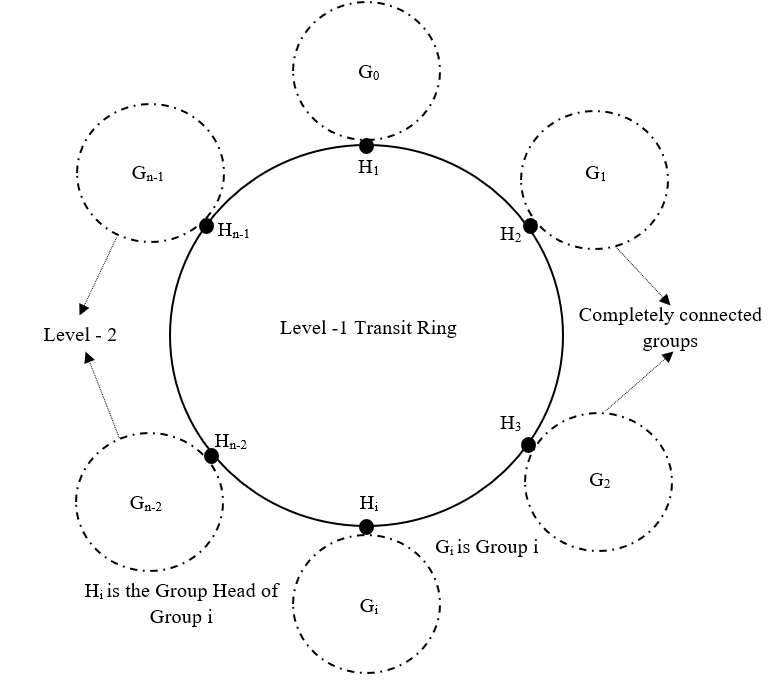} 
        \caption{A two-level RC based structured P2P architecture with n distinct resource types}
        \label{fig:figure1}
    \end{minipage}\hfill
    \begin{minipage}{0.7\textwidth}
        \centering
        \includegraphics[width=\textwidth]{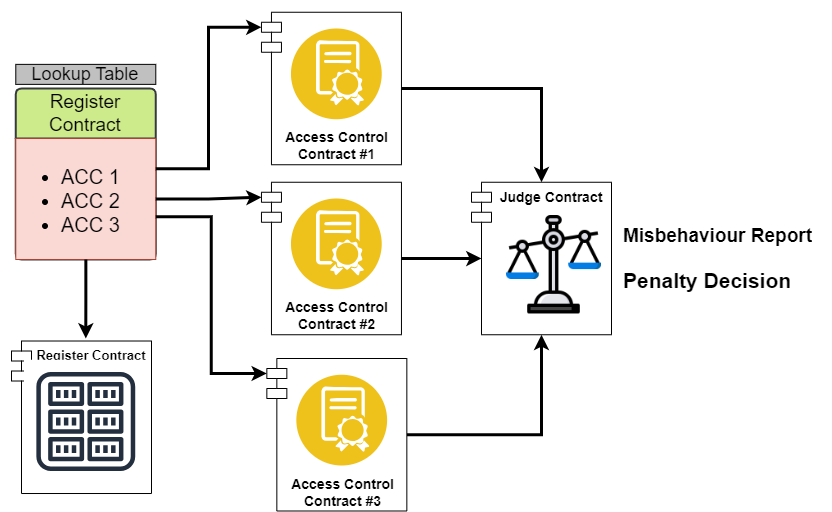} 
        \caption{Architecture of the Access Control System}
        \label{fig:arc}
    \end{minipage}
\end{figure}

\subsection{Access Control Contracts (ACCs)}
ACCs are the core components of the access control framework, implementing specific policies for subject-object pairs. These contracts manage various tasks within the P2P network such as viewing, editing, and managing the Global Resource Table (GRT), accessing local resources, and modifying group configurations. 

A dual validation mechanism is implemented to enhance security further. This mechanism includes:
\begin{itemize}[noitemsep, topsep=0pt]
    \item \textbf{Static Access Right Validation:} This layer checks predefined policies against access requests to ensure that all requests comply with granted permissions and are managed securely.
    \item \textbf{Dynamic Access Right Validation:} It evaluates the subject's behavior in real-time, adjusting access rights dynamically in response to emerging threats or abuses.
\end{itemize}
Additionally, each ACC maintains a misbehavior list detailing actions considered inappropriate, the timing of such actions, and the applied penalties. This record allows for dynamic adjustments to access permissions and maintains the discipline within the P2P system.
 \begin{enumerate}[noitemsep, topsep=0pt]
    \item \textbf{Storage}: All ACCs and their policies are stored on the Ethereum blockchain, ensuring immutability and transparency.
    \item \textbf{Decision Making}: Call acceptance and validation are decided by the collective consensus of the network nodes executing the smart contract code.
    \item \textbf{Failure vs Success}:
    \begin{itemize}[noitemsep, topsep=0pt]
        \item \textbf{Failure}: If the subject's request does not comply with predefined policies or fails adaptive validation due to suspicious behavior, the access request is denied, and the transaction is recorded on the blockchain.
        \item \textbf{Success}: If the subject's request meets all validation criteria, access is granted, and the transaction is recorded on the blockchain.
    \end{itemize}
    \item \textbf{Sample Tasks}: Examples of tasks managed by ACCs include:
    \begin{itemize}[noitemsep, topsep=0pt]
        \item \textbf{View GRT}: Allowing a subject to view the Global Resource Table.
        \item \textbf{Edit GRT}: Granting permissions to modify entries in the Global Resource Table.
        \item \textbf{Access Local Resources}: Enabling access to local resources based on the subject's role.
        \item \textbf{Modify Group Configurations}: Allowing changes to group member roles and resource allocations.
    \end{itemize}
\end{enumerate}
This table shows a representation of the permissions that different peers have. 

\begin{table}[ht]
\centering
\caption{Static Access permissions of different Subject}
\footnotesize
\begin{tabular}{|p{2cm}|p{2cm}|p{2cm}|p{2cm}|p{2cm}|p{2cm}|}
\hline
\textbf{Roles / Access} & \textbf{Global Resource Table} & \textbf{Local Resource Table} & \textbf{Malicious Group Head} & \textbf{Malicious Member} & \textbf{Communication Access} \\ \hline
Primary Group Head & full (view, edit, create, delete) & full (view, edit, create, delete) & full (view, edit, create, delete) & full (view, edit, create, delete) & other group head, own members \\ \hline
Secondary Group Head & view & view & view, edit & deny & own group head, own members \\ \hline
Regular Members & deny & view & deny & deny & own group head \\ \hline
\end{tabular}

\label{tab:access_control}
\end{table}

Table~\ref{tab:security_responses} presents a range of security policies designed to protect P2P systems by dynamically responding to different types of security violations. 

\begin{table}[ht]
\centering
\caption{Security event responses based on severity and impact}
\scriptsize
\setlength{\tabcolsep}{3pt} 
\begin{tabular}{|p{3cm}|p{1.5cm}|p{2cm}|p{2cm}|p{3cm}|}
\hline
\textbf{Security Event} & \textbf{Severity} & \textbf{CIA Impact} & \textbf{STRIDE Impact} & \textbf{Response} \\ \hline
Too many Access Attempts & High & Integrity & Elevation of Privilege & 24-hour ban \\ \hline
Data Tampering & High & Integrity & Tampering & Cancel access permanently \\ \hline
Unauthorized Access & Medium & Confidentiality & Information Disclosure & Temporary suspension \\ \hline
Disruption of Service & High & Availability & Denial of Service & 30 days suspension \\ \hline
Misrepresentation of Identity & Low & Accountability & Spoofing & Warning, penalties \\ \hline
\end{tabular}
\label{tab:security_responses}
\end{table}
\subsection{Judge Contract (JC)}

The JC judges the misbehavior of the subject and determines the corresponding penalty when receiving a potential misbehavior report from an ACC. After determining the penalty, the JC returns the decision to the ACC for further operation.
\begin{itemize}[noitemsep, topsep=0pt]
  \item \textbf{Object}: The peer who experienced the misbehavior. For example, a Primary Group Head (PGH) who noticed unauthorized access attempts to the Global Resource Table (GRT).
  \item \textbf{Misbehavior}: Detailed description of the misbehavior. For instance, multiple failed login attempts by a Secondary Group Head (SGH) trying to access restricted resources.
  \item \textbf{Time}: The specific time when the misbehavior occurred. An example would be a timestamp indicating when a Regular Member repeatedly tried to access Global Resource Table's Data.
  \item \textbf{Penalty}: The penalty imposed based on the severity of the misbehavior. For example, a temporary suspension of access rights for 24 hours for the Secondary Group Head due to multiple failed login attempts.
\end{itemize}

\subsection{Register Contract (RC)}
The Register Contract provides a centralized yet decentralized repository for all access control methods, ensuring that each method can be easily registered, updated, or deleted. It offers a system-level view of the entire network access control policies, which also provides dynamic reference data to different roles within the network, facilitating secure and efficient interactions.


The RC will also maintain a lookup table, which should look like this:

\begin{itemize}[noitemsep, topsep=0pt, parsep=0pt, partopsep=0pt]
  \item MethodName: the name of the method;
  \item Subject: the subject of the corresponding subject-object pair of the method;
  \item Object: the object of the corresponding subject-object pair of the method;
  \item ScName: the name of the corresponding smart contract implementing this method;
  \item Creator: the peer who created and deployed the contract;
  \item ScAddress: the address of the smart contract;
\end{itemize}

The RC provides the following main ABIs to manage these methods. 

\begin{itemize}[noitemsep, topsep=0pt, parsep=0pt, partopsep=0pt]
  \item \texttt{methodRegister()}: Verify a new method and register the information into the lookup table.
  \item \texttt{methodUpdate()}: Verification of an existing method that needs to be updated and updates the information, especially the fields of ScAddress and ABI.
  \item \texttt{methodDelete()}: Verifies the MethodName of a method and deletes the method from the lookup table.
  \item \texttt{getContract()}: Verifies the MethodName of a method and returns the address and ABIs of the contract (i.e., the ACCs and JC) of the method.
\end{itemize}

The RC is also engineered to dynamically provide reference data to different roles within the network based on their needs. For instance:

\begin{itemize}[noitemsep, topsep=0pt, parsep=0pt, partopsep=0pt]
  \item When a Primary Group Head acts as the subject and requests access to a resource or interaction, the RC supplies the reference data, such as the "Primary Group Head Role ACC," to facilitate the access or communication.
  \item Similarly, if a Secondary Group Head or Regular Member assumes the role of the subject, intending to connect with a resource or another entity, the RC provides them with their respective ACC references, like "Secondary Group Head Role ACC" or "Regular Members Role ACC," accordingly.
\end{itemize}

\begin{table}[ht]
\centering
\begin{tabular}{|p{3cm}|p{6cm}|}
\hline
\textbf{Subject Role} & \textbf{Reference Data Supplied by RC} \\ \hline
Primary Group Head & Primary Group Head Role ACC \\ \hline
Secondary Group Head & Secondary Group Head Role ACC \\ \hline
Regular Member & Regular Member Role ACC \\ \hline
\end{tabular}
\vspace{5pt}
\caption{Dynamic reference data supplied by the RC}
\label{tab:reference_data}
\end{table}

This dynamic reference mechanism ensures that each peer, depending on their role and the nature of the access request, is equipped with the appropriate smart contract references. It facilitates secure interaction within the network, ensuring that access controls are adhered to the predefined policies and permissions. 

In our blockchain-based system, the communication between primary group heads (PGHs) is efficiently managed through a series of interactions with smart contracts, specifically the Registry Contract and the Primary Group Head Role Access Control Contract (ACC). Detailed steps of the communication process are fully illustrated in the accompanying diagrams, which depict how access requests are handled, validated, and logged to maintain network security and integrity.

\begin{figure}[ht]
\centering
\includegraphics[width=0.9\textwidth]{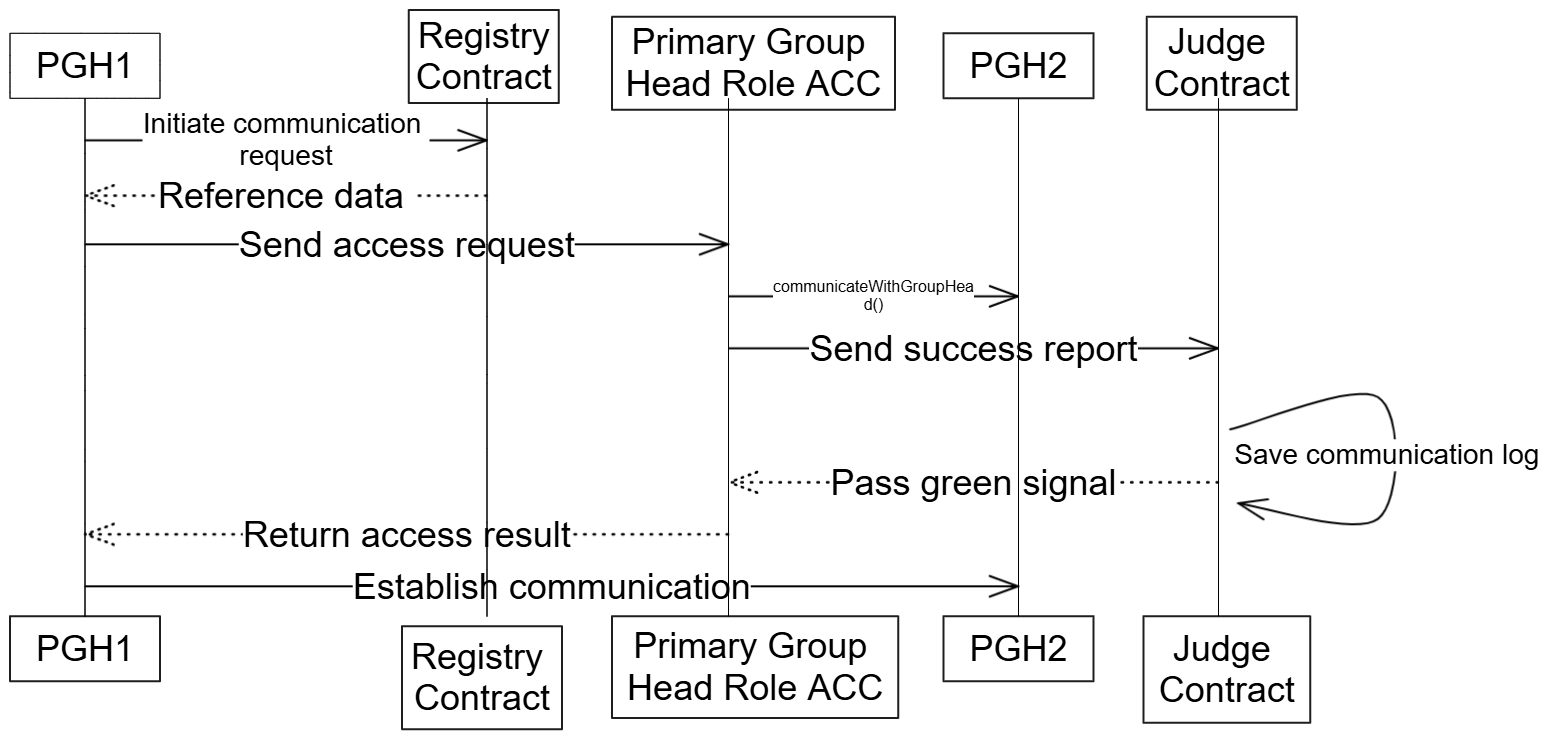}
\caption{Blockchain-based access control system architecture}
\label{fig:architecture}
\end{figure}

The blockchain-based access control system is designed to tightly regulate any updates or deletions made to Access Control Contracts (ACCs), ensuring that any modifications meet strict security and integrity standards. Whenever there are updates to the access control methods, they go through a thorough review process to make sure they're compatible with predefined roles and the overall system architecture. This helps maintain the security of the network and prevents any unauthorized access.

Two key contracts in our system are the Register Contract (RC) and the Judge Contract (JC). These contracts have important roles in overseeing the application of policies and the execution of roles within the network. They make sure that there are secure interactions among the Primary Group Heads (PGHs) and that the access control mechanisms are tailored to the specific needs of the network, ultimately improving operational efficiency.

The novel aspect of our work lies in the integration of blockchain technology with a dynamic access control mechanism that uses dual-validation. This mechanism is specifically designed for peer-to-peer (P2P) networks. By distributing the validation process across multiple nodes, the framework basically eliminate any single points of failure and continuously verify access requests and transactions. The system also adapts in real-time by adjusting permissions based on user behavior, allowing for immediate responses to any potential threats. By storing the ACCs and policies on the Ethereum blockchain, the system ensure that they cannot be changed or tampered with.

The dynamic reference data mechanism provided by the Register Contract adds role-specific access controls, which improves communication and overall efficiency. The architecture is built to handle high churn rates and the structural openness of P2P networks, making it scalable and flexible for networks of different sizes and structures.

In sum, the access control system, which is based on blockchain technology, along with the strict operational guidelines and the important roles played by the RC and JC, ensures a secure and efficient network. This approach not only improves communication and operational efficiency, but also maintains a high standard of security and integrity, which is crucial for the reliable operation of blockchain-based systems.

\subsection {Algorithm}
The access control logic defines roles and permissions for different group members, implementing functions for resource access and management. Below is the algorithm used to control access based on the role and resource table. The lines 2,7,23,19 correspond to the policies outlined in Table 1.

\begin{algorithm}
\caption{Access Control Logic}
\small 
\setlength{\textfloatsep}{5pt} 
\begin{algorithmic}[1]
\Function{AccessControl}{$resourceTable$}
    \If{$resourceTable == \text{Global Resource Table, Local Resource Table}$}
        \State \textbf{return} \text{Full access: view, edit, create, delete}
    \EndIf
    \State
    \Function{NormalPrimaryGH}{$resourceTable$}
        \If{$resourceTable == \text{Global Resource Table, Local Resource Table}$}
            \State \textbf{return} \text{view}
        \EndIf
    \EndFunction
    \State
    \Function{NormalSecondaryGH}{$resourceTable$}
        \If{$resourceTable == \text{Global Resource Table, Local Resource Table}$}
            \State \textbf{return} \text{view}
        \EndIf
    \EndFunction
    \State
    \Function{NormalRegularMember}{$resourceTable$}
        \If{$resourceTable == \text{Local Resource Table}$}
            \State \textbf{return} \text{view}
        \Else
            \State \textbf{return} \text{deny}
        \EndIf
    \EndFunction
\EndFunction
\end{algorithmic}
\end{algorithm}

The main function is \texttt{AccessControl}, which is called by the respective ACC based on the role and resource being accessed. For instance, when a Primary Group Head (PGH) attempts to access the resource table, the ACC invokes the \texttt{NormalPrimaryGH} function to determine the appropriate access level. This call is made dynamically by the smart contract handling the access request.

\section{Implementation} \label{Sec:implementation}

\textbf{Hardware Used}: We used HP Probook G4 440, which is equipped with an Intel(R) Core(TM) i7-7500U CPU, running at a frequency of 2.70GHz with 2 cores and 4 logical processors. Operating on Microsoft Windows 11 Pro (64 bit), the device has a memory capacity of 15.9 GB. For storage, it offers a 931.51 GB HDD and a 223.57 GB SSD.

The various contracts implemented within our blockchain-based access control system, each tailored to specific roles and their corresponding access levels as depicted in Table 1 of subsection 4.1.

\textbf{BaseAccessControl Contract}: This foundational contract holds basic resource management and internal communication. Serving as a template, it allows for modular extensions specific to each role, laying out a framework where general permissions are specified further.

\textbf{PrimaryGroupHeadRoleACC Contract}: Extending from \texttt{BaseAccessControl}, this contract provides Primary Group Heads with comprehensive resource management rights. It effectively translates the conceptual ``Full access'' into practical abilities to view, edit, create, and delete resources both locally and across the network. 

\textbf{SecondaryGroupHeadRoleACC Contract}: Tailored for Secondary Group Heads, this contract confines its functionalities to viewing resources only, aligning with the prescribed limited access. It enforces strict adherence to the hierarchical access control by prohibiting any modifications to resources.

\textbf{RegularMembersRoleACC Contract}: This contract limits Regular Members to accessing local resources only, consistent with the stringent access controls outlined by the algorithm. 

\begin{figure}[htbp]
    \centering
    \includegraphics[scale=0.3]{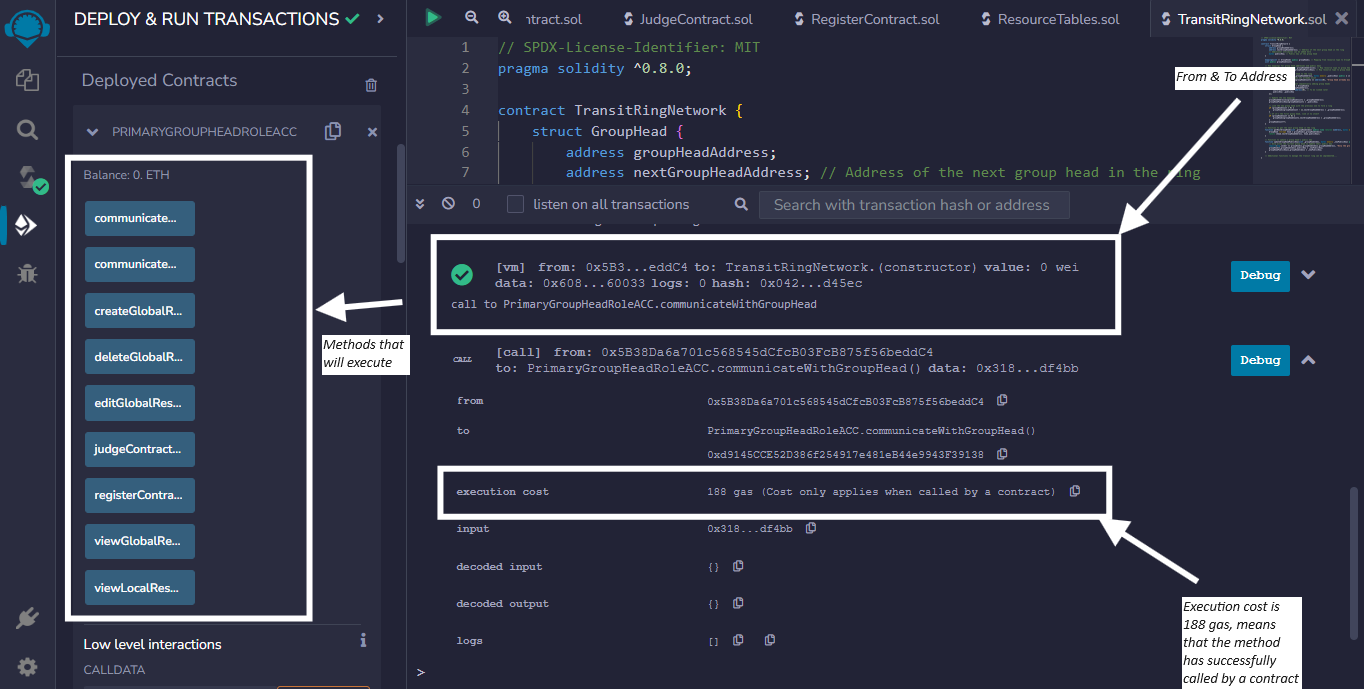} 
    \caption{Description of the PrimaryGroupHeadRoleACC contract and the associated transaction.}
    \label{fig:acc}
\end{figure}

At the bottom, we have the input data, followed by the decoded inputs, outputs, and logs from the function, indicating that the execution was successful. As shown in Figure 4, this process confirms the successful execution.

\section{Result and Discussion} \label{Sec:result}

The blockchain-based access control framework for P2P networks applied in this case has resulted in advanced levels of network security and effectiveness as we can see in the comparison of system performance Before and after the Implementation below:

\begin{table}[htbp]
\centering
\begin{tabular}{|p{3.5cm}|p{4.5cm}|p{4.5cm}|}
\hline
\textbf{Category} & \textbf{Before Implementation} & \textbf{After Implementation} \\
\hline
Response Times & Slow & Improved \\
\hline
Unauthorized Access & High rates & Reduced \\
\hline
Access Management Transparency & Lack of transparency & Enhanced \\
\hline
\end{tabular}
\label{tab:performance_comparison}
\end{table}

\section{Conclusion} \label{Sec:conclusion}

This paper presents a blockchain-based access control framework for high-security peer-to-peer (P2P) networks, specifically in IoT and Web 3.0 environments. Tested in an RC-based P2P network, it leverages Ethereum smart contracts for immutable, transparent, decentralized security management, reducing vulnerabilities. The novelty of the work lies in integrating blockchain with dynamic, dual-validation access control, enhancing security by distributing validation across nodes, verifying access requests, and eliminating single points of failure. Real-time adaptability ensures immediate threat response, while immutable Ethereum records build trust. 

Future work includes developing additional Access Control Contracts (ACCs) to handle other functions and roles within the network, further enhancing the granularity and security of our access control system. 

\bibliographystyle{plain}
\bibliography{MAIN}

\end{document}